\documentclass[journal,article,accept,oneauthor,pdftex,12pt,a4paper]{mdpi} 

\lastpage{\pageref*{LastPage}}
\doinum{10.3390/e1705xxxx}
\pubvolume{17}
\pubyear{2015}
\history{Received: 9 April 2015 / Accepted: 4 May 2015 / Published: xx May 2015}

\usepackage{graphicx,soul}

\usepackage{color}
\definecolor{dark-green}{RGB}{0,100,0}
  \usepackage{amsmath}
  \usepackage{bm}
  \newcommand{\Espaco}{\rule[-5mm]{0mm}{13mm}} 
  \newcommand{\de}[1]{\left(#1\right)}

\Title{Conceptual Inadequacy of the Shore and Johnson Axioms for Wide Classes of Complex Systems} 

\Author{Constantino Tsallis $^{1,2}$}

\address{%
$^{1}$ Centro Brasileiro de Pesquisas F\'{\i}sicas and 
National Institute of Science and Technology for Complex Systems, 
Rua Xavier Sigaud 150, 22290-180, Rio de Janeiro - RJ, Brazil; E-Mail: tsallis@cbpf.br \\
$^{2}$ Santa Fe Institute, 1399 Hyde Park Road, Santa Fe, New Mexico 87501, NM,
USA}

\abstract{It is by now well known that the Boltzmann-Gibbs-von Neumann-Shannon logarithmic entropic functional ($S_{BG}$) is inadequate for wide classes of strongly correlated systems: see for instance the 2001 Brukner and Zeilinger's {\it  Conceptual inadequacy of the Shannon information in quantum measurements},
 among many other systems exhibiting various forms of complexity. On the other hand, the Shannon and Khinchin axioms uniquely mandate the BG  form $S_{BG}=-k\sum_i p_i \ln p_i$; the  Shore and Johnson axioms follow the same path. Many natural, artificial and social systems have been satisfactorily approached with nonadditive entropies such as the $S_q=k \frac{1-\sum_i p_i^q}{q-1}$ one ($q \in {\cal R}; \,S_1=S_{BG}$), basis of nonextensive statistical mechanics. Consistently, the Shannon 1948 and Khinchine 1953 uniqueness theorems have already been generalized in the literature, by Santos 1997 and Abe 2000 respectively, in order to uniquely mandate $S_q$. We argue here that the same remains to be done with the Shore and Johnson 1980  axioms. We arrive to this conclusion by analyzing specific classes of strongly correlated complex systems that await such generalization.
}


\begin{document}

In classical mechanics we are allowed to simultaneously measure a coordinate and its conjugate momentum. Not so in quantum mechanics, where these two variables appear as correlated in a well known subtle manner. Planck constant $h$ precisely plays the role of characterizing this correlation. It is in this sense that the formal limit $h \to 0$ makes quantum mechanics to recover the classical one. Similarly, space and time play independent roles in classical mechanics. Not so in special relativity, where they mutually influence each other through the Lorentz transformation. The vacuum light velocity $c$ precisely plays the role of characterizing this nontrivial mutual influence. It is in this sense that the formal limit $1/c \to 0$ makes relativistic mechanics to recover the classical one. In the theory of probabilities and statistical mechanics, a totally analogous situation occurs.  Indeed, if systems $A$ (with probabilities $\{ p_i^A\},\, i=1,2,\dots,W_A$) and $B$ (with probabilities $\{ p_j^B\},\,j=1,2, \dots,W_B$) are {\it probabilistically independent}, the joint probabilities of the system $(A+B)$ are simply given by $p_{ij}^{A+B}=p_i^A p_j^B$.  If, however, correlations exist between $A$ and $B$, this lack of independence can be characterized by some parameter(s) (from now on noted $\kappa$), which can always be defined in such a way that independence is recovered when $\kappa$ vanishes. In such situation we have that $p_{ij}^{A+B}=f_{ij}(p_i^A,p_j^B; \kappa)$ where  $f_{ij}$ is a function typically satisfying $f_{ij}(x,y; 0)=xy$. All these rather elementary remarks apply to all kinds of approaches in theoretical physics and elsewhere: related arguments can be seen, for instance, in \cite{BruknerZeilinger2001}.

In a recent paper \cite{Presseetal2013} (see also \cite{Presse2014}), Press\'e {\em et al.}, claim that they proved that {\it nonadditive} entropies like those yielding nonextensive statistical mechanics~\cite{Tsallis1988,GellMannTsallis2004,Tsallis2009a,Tsallis2009b} (as well as many others such as the Anteneodo-Plastino entropy, Curado entropy, Borges-Roditi entropy, Kaniadakis entropy, among others) are logically inadmissible because they would introduce biases that are not warranted by the data. They base their arguments on the Shore and Johnson (SJ) set of axioms. This set is consistent with the Boltzmann-Gibbs (BG) entropy, which, as well known, is {\it additive} in the Penrose sense \cite{Penrose1970} ({\em i.e.}, additive with regard to {\it probabilistically independent} subsystems).  

Here we emphasize that nonadditive entropies emerge from strong correlations (among the random variables of the system) which are definitively out of the SJ hypothesis. 
Indeed, the interesting SJ axiomatic framework addresses essentially systems for which it is possible to provide information about their subsystems without 
providing information about the interactions between the subsystems, which clearly is {\it not} the case where nonadditive entropies are to be used.  Let us be more specific. The SJ axioms focus on systems whose total number $W(N)$ of microscopic possibilities exponentially increases with $N$ in the $N\to\infty$ limit, {\em i.e.}, when $W(N) \sim \mu^N \,(\mu >1)$ (hereafter referred to as the {\it independent} or {\it exponential class}), or equivalently when $W(N+1) \sim W(N)\times \mu$, where $\mu$ essentially corresponds to the nonvanishing Lebesgue measure of the one-particle phase space. The entropy to be used for such systems clearly is (in its discrete version) the BG one, $S_{BG}=k \sum_{i=1}^W p_i \ln \frac{1}{p_i}$ (with $\sum_{i=1}^W p_i=1$), which scales with $N$ like its equal-probability particular instance, namely $S_{BG}=k \ln W$. Indeed, for the so called {\it exponential  class} (which includes strict independence as the particular case $W(N) = \mu^N$, with $\mu >1)$), we have $S(N)=k\ln W(N) \sim N$, thus providing an {\it extensive} entropy as thermodynamically required.

In relevant contrast with the above case, strong correlations between the $N$ elements do exist in great variety of natural, artificial and social systems. For example, it might be $W(N) \sim N^\rho \,(\rho >0)$ (referred to as the {\it power-law class}), hence $W(N+1) \sim W(N) (\frac{N+1}{N})^\rho \sim W(N)(1+\frac{\rho}{N})$. The entropy corresponding to this case is $S_q=k \frac{1-\sum_{i=1}^W p_i^q}{q-1}=k\sum_{i=1}^W p_i \ln_q \frac{1}{p_i}=   -k  \sum_{i=1}^W p_i^q \ln_q p_i = -k\sum_{i=1}^W p_i  \ln_{2-q} p_i $ with $\ln_q z \equiv \frac{z^{1-q}-1}{1-q} \, (z>0; \, \ln_1 z=\ln z)$ and $q=1-\frac{1}{\rho}$. Indeed, the extremal $S_q$ (occurring for equal probabilities) is given by $S_q=k\ln_q W$, and, for that specific value of $q$, we verify that $S_{q=1-1/\rho}(N) \propto N$, thus once again providing an {\it extensive} entropy as thermodynamically required. Another example of strong correlations between the $N$ elements of the system occurs when $W(N) \sim \nu^{N^\gamma} \,(\nu>1;\, 0<\gamma <1)$ (referred to as the {\it stretched-exponential class}), hence $W(N+1) \sim W(N) \frac{\nu^{(N+1)^\gamma}}{\nu^{N^\gamma}} \sim W(N) (1+ \frac{\gamma \ln \nu}{N^{1-\gamma}})$.
An entropy corresponding to this case is $S_\delta=k\sum_{i=1}^W p_i (\ln \frac{1}{p_i})^\delta$ with $\delta=1/\gamma$ (footnote in page 69 of \cite{Tsallis2009a}; see also \cite{Ubriaco2009}). Indeed, the extremal $S_\delta$ (occurring for equal probabilities) is given by $S_\delta=k(\ln W)^\delta$, and, for that specific value of $\delta$, we verify that $S_{\delta=1/\gamma}(N) \propto N$, thus once again providing an {\it extensive} entropy as thermodynamically required. Notice that, in the limit $N\to\infty$, $1 \ll N^\rho \ll \nu^{N^\gamma} \ll \mu^N$, which means that the phase-space Lebesgue measure corresponding to the power-law and stretched-exponential classes vanishes, whereas that corresponding to the exponential class remains finite. This is the deep reason why, in the presence of strong correlations (e.g., the power-law and the stretched-exponential classes), the BG entropy is generically not physically appropriate anymore since it violates the thermodynamical requirement of entropic extensivity (the reasons why entropy is in all cases required to be thermodynamically extensive remains outside the scope of the present paper; the interested reader can however refer to \cite{TsallisCirto2013,KomatsuKimura2013,Komatsu2014,KomatsuKimura2014,RuizTsallis2013}). The above cases are summarized in Table \ref{table1}. We see therefore that the SJ set of axioms, {\it demanding, as they do, system independence}, are {\it not applicable unless we have indisputable reasons to believe that the data that we are facing correspond to a case belonging to the exponential class}, and by no means correspond to strongly correlated cases such as those belonging to the power-law or stretched-exponential classes, or even others \cite{Tempesta2011, HanelThurner2011a,HanelThurner2011b}. Even if we were not particularly interested in thermodynamics, it is clear that the scenario of system independence must be assumed for the data in order to justify the use of the SJ framework. If we may insist, the actual data must be minimally consistent with a scenario of system independence; if they are not, the whole inference procedure needs by all means to follow a different path. 
This central point is missed in \cite{Presseetal2013}. Incidentally, there are in  \cite{Presseetal2013} a few more inadvertences or inadequacies, the most prominent of which we highlight now.

\begin{table}[h]
\centering
\begin{tabular}{|c|c|c|c|}   \cline{2-4} 
\multicolumn{1}{c|}{}                                &\multicolumn{3}{c|}{\rule[-3mm]{0mm}{9mm}\footnotesize{\textbf{ENTROPY}} } \\  \hline
\rule[-1mm]{0mm}{7mm}$\bm{W\de{N}}$                  &                   $S_{BG}$                                 & $S_{q}$                                                       & $S_{\delta}$                                                 \\
\rule[-2mm]{0mm}{7mm}$\bm{\de{N \to \infty}}$        &                                                            & $\de{q\neq1}$                                                 & $\de{\delta\neq1}$                                           \\
\rule[-3mm]{0mm}{7mm}                                & \textcolor{dark-green}{\textbf{\footnotesize{(ADDITIVE)}}} & \textbf{\textcolor{dark-green}{\footnotesize{(NONADDITIVE)}}} &\textbf{\textcolor{dark-green}{\footnotesize{(NONADDITIVE)}}} \\ \hline 
\Espaco$\displaystyle{{\sim\mu^{N}} \atop \de{\mu > 1} }$  &\textbf{\textcolor{blue}{\footnotesize{EXTENSIVE}}}   & \textbf{\textcolor{red}{\footnotesize{NONEXTENSIVE}}}  &\textbf{\textcolor{red}{\footnotesize{NONEXTENSIVE}}}              \\ \hline
\Espaco$\displaystyle{\sim N^{\rho} \atop \de{\rho > 0} }$ &\textbf{\textcolor{red}{\footnotesize{NONEXTENSIVE}}} & \textbf{\textcolor{blue}{\footnotesize{EXTENSIVE}}}    &\textbf{\textcolor{red}{\footnotesize{NONEXTENSIVE}}}              \\
\rule[-2mm]{0mm}{0mm}                                      &                                                      & $\de{q = 1- 1/\rho}$                                   &                                                                   \\ \hline
\Espaco$\displaystyle{\sim\nu^{ N^\gamma } \atop (\nu>1; }$&\textbf{\textcolor{red}{\footnotesize{NONEXTENSIVE}}} & \textbf{\textcolor{red}{\footnotesize{NONEXTENSIVE}}}  & \textbf{\textcolor{blue}{\footnotesize{EXTENSIVE}}}               \\
\rule[-2mm]{0mm}{0mm}$0<\gamma<1)$                         &                                                      &                                                        & $\de{\delta = 1/\gamma}$                                          \\ \hline
\end{tabular}
\caption{Additive and nonadditive entropic functionals and classes of systems for which the entropy is extensive. $W(N)$ is the number of admissible microscopic configurations of a system with $N$ elements; only configurations with nonvanishing occurrence probability are considered admissible.}
\label{table1}
\end{table}

(a) {\it On the nature of the index $q$:}
It is emphasized in \cite{Presseetal2013} that $q$ would be obtained {\it not} from first principles but only as a {\it fitting parameter}. To support such a viewpoint, the authors quote a nontechnical article published over a decade ago (their Ref. [25] in \cite{Presseetal2013}). They seem to be unaware of the many examples where $q$ has in fact been obtained from (dynamical) first principles. Let us quote here some of them, among many others: (i) The value of $q$ (both for the sensitivity to the initial conditions and the entropy production per unit time) at the Feingenbaum point of the logistic map is \cite{LyraTsallis1998,BaldovinRobledo2004} $q=1-\frac{\ln 2}{\ln \alpha_F}=0.2444877013....  $ (1018 exact digits are presently known: see for instance \cite{Tsallis2009a}; $\alpha_F$ is the Feigenbaum universal constant); 
(ii) The value of $q$ for the $q$-Gaussian distribution of velocities of cold atoms in optical lattices is given \cite{Lutz2003,DouglasBergaminiRenzoni2006,LutzRenzoni2013} by $q=1+44E_R /U_0$, where $E_R$ is the recoil energy and $U_0$ the potential depth;
(iii) The value of $q$ in order to have an extensive block entropy in the class of one-dimensional systems at a quantum critical point characterized by the central charge $c$ is given by $q= \frac{\sqrt{9+c^2} -3}{c} \, (c>0;$ notice that $c \to\infty$ yields $q \to 1)$ \cite{CarusoTsallis2008}; 
(iv) The index $q$ for the  stationary-state distribution in space (and, in fact, also in momenta) in overdamped systems of the type-II superconductors in the presence of an external confining potential is given \cite{AndradeSilvaMoreiraNobreCurado2010} by a $q$-exponential of the potential with $q=0$, which, within the variational framework which uses linear constraints, corresponds to an entropic index $q=2-0$. 

However, even if it is definitively wrong that $q$ cannot be derived from first principles, it is certainly true that, in the literature, it does frequently play the role of a fitting parameter. This is a natural consequence of the simple fact that the calculation of $q$ from first principles demands the exact knowledge of the microscopic dynamics, which is very rarely available (and, even when it is available, it frequently involves mathematically intractable calculations). This has no deeper epistemological significance than the well known fact that the precise orbits of the planets of our planetary system are not in practice obtained from first principles but rather from astronomical observations {\it and fittings}. This by no means implies that Newtonian mechanics is not a first-principle theory, but it just exhibits that its full calculation is virtually impossible because the exact knowledge of the initial conditions of all the masses of the planetary system is required. In spite of that gigantic difficulty, classical mechanics is nevertheless capable of easily establishing the (approximate) {\it form} of all those orbits, namely the Keplerian elliptic form. Similarly, even when $q$ is not analytically accessible, basic properties for wide classes of complex systems can be satisfactorily fitted with the $q$-exponential and $q$-Gaussian forms that are implied by the extremization  of the $S_q$ entropy.\\

(b) {\it On nonindependent probabilities:}
We read in \cite{Presseetal2013} that  ``the Tsallis entropy can only be justified if events $i$ and $j$ were to have the following joint probability,
$p_{ij}^{q-1}=(u_i^{q-1}+v_j^{q-1}-1)$'', and ``We apply SJ's approach to derive what joint probability for states of two systems would be required to justify the form of the Tsallis entropy''. In the notation currently used in $q$-entropies and nonextensive statistical mechanics, these statements correspond to $1/p_{ij} = (1/u_i) \otimes_q (1/v_j)$, or alternatively to  $p_{ij} = u_i \otimes_{2-q} v_j$ , where $x \otimes_q y \equiv [x^{1-q} + x^{1-q}-1]^{1/(1-q)} \,(x \otimes_1 y=xy)$ is referred to as the {\it $q$-product} \cite{NivanenLeMehauteWang2003,Borges2004}. 
There is in \cite{Presseetal2013} some degree of confusion with respect to this point. Let us be precise. If we have the particular instance of equal probabilities, then $S_q=k \ln_q W$, hence indeed we have the extensive-like equality $S_q(W_u \otimes_q W_v) = S_q(W_u) + S_q(W_v)$, to be contrasted with the nonadditive expression $\frac{S_q(W_u \times W_v)}{k}= \frac{S_q(W_u)}{k} + \frac{S_q(W_v)}{k} + (1-q) \frac{S_q(W_u)}{k} \frac{S_q(W_v)}{k}$, $W_u$ and $W_v$ being the total number of states of the systems with probabilities $\{u_i\}$ and $\{v_j\}$ respectively. But, for the generic case ({\em i.e.}, when we have {\it not necessarily} equal probabilities), we do {\it not} have the simple expression indicated in \cite{Presseetal2013} for the composition of $q$-entropies. Indeed, $S_q(\{u_i \otimes_{2-q} v_j\})= -k \sum_{ij} (u_i \otimes_{2-q} v_j) \ln_{2-q} (u_i \otimes_{2-q} v_j) =  -k \sum_{ij} (u_i \otimes_{2-q} v_j) (\ln_{2-q} u_i + \ln_{2-q} v_j)
\ne  -k \sum_{ij} u_i v_j (\ln_{2-q} u_i + \ln_{2-q} v_j) = -k\sum_{i=1}^W u_i \ln_{2-q} u_i - k\sum_{j=1}^W v_j \ln_{2-q} v_j = S_q(\{u_i\}) + S_q(\{v_j\})$. This is an interesting and nontrivial consequence of this class of correlations  (curiously enough, referred in \cite{Presseetal2013} to as ``spurious correlations'')
between probabilistic events.\\

(c) {\it On unconventional averages:}
We read in \cite{Presseetal2013} that  ``Furthermore, unconventional averages must often be used to constrain nonadditive entropies (their Refs. [28--33]) to assure the convexity of those functions if they are to be used to infer a unique model.'' This claim appears to be kind of unclear since, for example, the {\it nonadditive} entropic functional $S_q$ is itself concave for all $q>0$ (in contrast, for instance, with the {\it additive} Renyi entropic functional, which is concave only for $0 < q \le 1$). When we want to go one step further, and produce a statistical mechanics, then constraints must indeed be added, and they naturally play a relevant role. From an information-theory standpoint, constraints must be simple and robust, such as mean values and widths. These quantities are typically taken to be $\langle x \rangle$ and $\langle x^2 \rangle$, since they simply characterize the location and the width (given by $\sqrt{\langle x^2 \rangle - \langle x\rangle^2}$) of the distribution of probabilities $p(x)$ of the random variable noted $x$. These quantities are mathematically admissible {\it as long as they are finite} (which typically occurs if $p(x)$ decays quickly enough, for example exponentially quick, at large values of $|x|$). When we have very-long-tailed power-laws these quantities diverge, and we {\it must} therefore replace the conventional constraints  by mathematically appropriate ones. For example, the mean value $\langle x \rangle$ for the $q$-exponential distribution $p(x) \propto e_q^{- \beta x}$ (with $\beta >0$ and $x \ge 0$; $e_q^z \equiv [1+(1-q)z]^{1/(1-q)}$, and $e_1^z=e^z$) diverges for $q \ge 3/2$, whereas appropriately $q$-generalized mean values remain {\it finite} as long as the distribution $p(x)$ remains normalizable (in this example, for $q<2$). Another example: if we had $p(x) \propto e_q^{-\beta x^2} \,(\beta >0; -\infty < x< \infty)$, then $\langle x^2 \rangle$ diverges for $q \ge 5/3$, whereas the appropriately $q$-generalized variance remains {\it finite} as long as the distribution $p(x)$ remains normalizable (in this example, for $q<3$). These important mathematical issues are illustrated in \cite{ThistletonMarshNelsonTsallis2007} and analytically discussed in \cite{TsallisPlastinoAlvarez2009}; see also \cite{FerriMartinezPlastino2005} for transformations associated with constraints of different kinds within nonextensive statistical mechanics. 
In the case of power-laws, the authors of \cite{Presseetal2013} offer as alternative  to use $S_{BG}$ with logarithmic constraints: ``We conclude by adding that it is possible to infer power laws within a principle of maximizing the BG entropy by constraining just one average: Mandelbrot (their Ref. [35]) showed this by invoking logarithmic constraints $\langle \ln x \rangle$ [to avoid confusion, we have noted $x$, and not $k$ the random variable]''. If we do this, we obtain $p(x) \propto 1/x^\eta \,(\eta \in {\cal R}; \, 0<x < \infty)$, which is {\it not normalizable for any value of $\eta$}. In fact, given any admissible entropy and any distribution of probabilities, it is always possible to find constraints such that the extremization of that entropy yields precisely that distribution.  Clearly, it is not in this manner that theory of information is to be used. 
Another most relevant aspect which also justifies the use, within a variational framework, of both additive and nonadditive entropies is the following. If we consider the standard (homogeneous and linear) Fokker-Planck equation in the presence of an external confining potential $V(x)$, we obtain for its unique stationary state $p(x) \propto e^{- \beta V(x)}$, which {\it precisely coincides} with the distribution which uniquely optimizes $S_{BG}$ with fixed $\langle V(x) \rangle$. If we consider now quite general inhomogeneous and/or nonlinear Fokker-Planck equations still in the presence of the confining potential $V(x)$, its unique stationary state once again {\it precisely coincides} with the distribution extremizing specific generalized entropies satisfying the $H$-theorem, and whose functional form is mandated by the specific inhomogeneity and/or nonlinearity \cite{SchwammleNobreCurado2007,SchwammleCuradoNobre2007,SchwammleCuradoNobre2009,RibeiroNobreCurado2011,MarizTsallis2012,RibeiroTsallisNobre2013}. \\

Last but not least, a considerable amount of predictions, verifications and applications of nonadditive entropies are today available in the literature which are useful in theoretical, experimental, observational and computational approaches of a wide variety of systems.
Among many others, we may illustrate the existing bibliography \cite{biblio} with: 
(i) The velocities of cold atoms in dissipative optical lattices~\cite{DouglasBergaminiRenzoni2006};  
(ii) The velocities of particles in quasi-two dimensional dusty plasma~\cite{liugoreeprl08};
(iii) Single ions in radio frequency traps interacting with a classical buffer gas~\cite{devoe}; 
(iv) The relaxation curves of RKKY spin glasses, like CuMn and AuFe~\cite{pickup}; 
(v) Hadronic transverse momenta distributions at LHC experiments~\cite{WongWilk2013};
(vi) Long-ranged-interacting many-body classical Hamiltonians \cite{CirtoAssisTsallis2014,ChristodoulidiTsallisBountis2014};
(vii) Nonlinear generalizations of the Schroedinger, Klein-Gordon and Dirac equations through $q$-plane waves \cite{NobreMonteiroTsallis2011,Borges1998};
(viii) The black-hole entropy and the so called area-law in quantum systems \cite{TsallisCirto2013,KomatsuKimura2013,Komatsu2014,KomatsuKimura2014};
(ix) The (area-preserving) standard map \cite{TirnakliBorges2015}.
Also $q$-generalized forms, still under progress, of the Central Limit Theorem have emerged from this theory \cite{CLT1,CLT2}. 

For better understanding the domain of validity of the SJ {\it System Independence Axiom} (Axiom III of Ref. [14] of \cite{Presseetal2013}), let us focus on a specific case.
Consider, for example, a one-dimensional system at zero temperature, where it displays a quantum critical point, say an Ising ferromagnetic chain in the presence of a transverse magnetic field at its critical value. Let us assume that the chain has $N$ first-neighbor-connected spins, and let us focus on a subsystem of it made of $L$ successive spins, see for instance  \cite{CarusoTsallis2008}. Let us note $\rho_L \equiv Tr_{N-L} \rho_N$
the density matrix of the subset, which is obtained by tracing over $(N-L)$ spins the density matrix $\rho_N$ of the total chain. We define the block $q$-entropy as follows: $S_q(L,N) = k\frac{1-Tr (\rho_L)^q}{q-1}$, $S_1(L,N) = -kTr \rho_L \ln \rho_L$ being of course the block Boltzmann-Gibbs-von Neumann entropy. 
The entropy of the total chain is given by $S_q(N,N)$. For the strongly quantum-entangled state that we are considering here the system is in a {\it pure state} ({\em i.e.}, $Tr \rho_N^2 =1$), therefore $S_q(N,N)$ vanishes for all $N$ and all $q$, in particular for $q=1$. Consequently the asymptotic entropy per particle $\lim_{N\to\infty} \frac{S_q(N)}{N}$ also vanishes, $\forall q$. 
In remarkable contrast, the block of $L$ spins is in a very different state, namely a {\it statistical mixture} ({\em i.e.}, $Tr \rho_L^2 <1$). Consequently, the block entropy per particle has a very different behavior than that of the full entropy per particle. More precisely, $\lim_{L\to\infty} \frac{S_q(L,\infty)}{L}$ {\it vanishes} for $q>q^*$, {\it diverges} for $q<q*$, and is {\it finite} for $q=q^*$, where $q^* <1$, and $S_q(L,\infty) \equiv \lim_{N\to\infty} S_q(L,N)$ \cite{CarusoTsallis2008}. 
All these somewhat counterintuitive facts come from the fact that, if we divide the full chain into say an $L$-block and an $(N-L)$-block, it is {\it not} possible to provide informations on each of the two blocks that would {\it not} interact, in the sense that doing measurements on one of them would not affect the other one. 
This fact does not accommodate with the SJ set of axioms (particularly the System Independence Axiom). The same happens for the power-law and stretched-exponential classes discussed earlier in this paper.

The well known set of axioms of Shannon \cite{Shannon1948} and of Khinchin \cite{Khinchin1953} mandate a unique entropic functional, namely $S_{BG}$. They have both been $q$-generalized (in \cite{Santos1997} and \cite{Abe2000} respectively), and the unique entropic functional that is then mandated is $S_q$. Let us conclude by mentioning that we believe that, similarly, the Shore-Johnson axioms might be generalizable in order to cover $S_q$ or even other nonadditive entropies. Undoubtedly, such a generalization would definitively be welcome.

\acknowledgments{Acknowledgments}
I thank L.J.L. Cirto, E.M.F. Curado, M. Jauregui, F.D. Nobre and A.R. Plastino for very fruitful conversations. 
Partial financial support from the Brazilian agencies CNPq and FAPERJ and the John Templeton Foundation is gratefully acknowledged. 

\conflictofinterests{Conflicts of Interest}

The author declares no conflict of interest. 


\end{document}